# Far detuned mid-infrared frequency conversion via normal dispersion modulation instability in chalcogenide microwire

Thomas Godin[1,*], Yves Combes[1], Raja Ahmad[2], Martin Rochette[2], Thibaut Sylvestre[1], and John M. Dudley[1]

[1]Institut FEMTO-ST, UMR 6174 CNRS-Université de Franche-Comté, Besançon, France
[2]Department of Electrical and Computer Engineering, McGill University, Montreal (QC) H3A 2A7, Canada
*Corresponding author: thomas.godin@femto-st.fr

We report the observation of modulation instability in the mid-infrared spectral region by pumping a hybrid polymer-chalcogenide optical microwire with a femtosecond optical parametric oscillator operating at 2.6 µm. It is further shown that this modulation instability occurs in the normal dispersion regime through negative fourth-order dispersion and leads to far detuned parametric frequency conversion at 2 µm and 3.5 µm, despite the presence of a strong absorption band around 2.8 µm. Stochastic nonlinear Schrödinger equation simulations of mid-infrared modulation instability are in excellent agreement with experiments.

Chalcogenide glasses have been identified as extremely attractive materials for applications in mid-infrared (mid-IR) photonics. Their advantages over silica-based and soft glasses include broad IR transparency and enhanced material nonlinearity [1,2]. As a result, there has been extensive interest in the use of chalcogenide glasses such as $As_2Se_3$ or $As_2S_3$ for applications in mid-IR broadband supercontinuum (SC) generation [3-5], stimulated Raman and Brillouin scattering [6-8].

Another fundamental nonlinear optical process is spontaneous modulation instability (MI), the parametric amplification of low amplitude noise on a pump wave associated with the growth of symmetric spectral sidebands about an incident pump [9]. MI plays a central role in long pulse SC generation [10], and has recently been the subject of much renewed interest in the context of understanding extreme instability processes in optics [11-13]. Significantly, although it was widely considered that phasematching spontaneous MI required pumping in the anomalous dispersion regime, scalar MI can also be phasematched with a normal dispersion regime pump provided that the fiber used possesses a suitable higher-order group velocity dispersion profile. This has been previously shown in specialty dispersion-shifted fiber and photonic crystal fiber (PCF), where widely spaced narrow-band frequency conversion up to several tens of THz has been observed [14,15].

In this Letter, we show that the same process of broadband frequency conversion due to normal dispersion MI can also be observed in specially tapered highly nonlinear hybrid $As_2Se_3$ polymer tapered microwires [16-20] where the dispersion profile allows phasematched frequency conversion via fourth order dispersion. Our results show far-detuned parametric frequency conversion at 2 µm and 3.5 µm by pumping the chalcogenide microwire in the normal dispersion regime at 2.62 µm. Calculated phasematching conditions and numerical simulations including the variation of nonlinearity and dispersion along the microwire are shown to be in very good agreement with experiment. The measured 30 THz frequency shift is among the largest reported using normal-dispersion MI in the mid-IR, although frequency shifts larger than 100 THz have been obtained under other experimental conditions [21,22].

The experimental setup is shown schematically in Fig. 1(a). For the pump source, we used the idler output of an optical parametric oscillator (OPO - Coherent Chameleon System) operating at a 80 MHz repetition rate. The OPO idler was tuned to 2.62 µm and the pulse duration was measured at ~600 fs using an intensity autocorrelator. A ZnSe-based mid-IR focusing objective was then used to couple IR light into the chalcogenide microwire and output spectra were recorded using a mid-IR-compatible large-core fiber coupled to a Fourier-transform infrared (FTIR) spectrometer (Arcoptix FTIR-Rocket) with a 2.7 nm resolution in the 2-6 µm wavelength range.

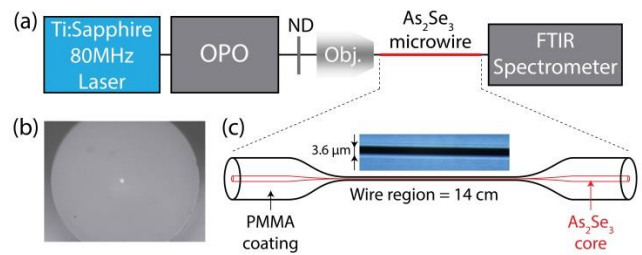

Fig. 1 (a) Experimental setup. OPO: Optical Parametric Oscillator. ND: Neutral Density Filter. Obj.: ZnSe mid-IR objective lens. (b) Input facet of the hybrid microwire. (c) Schematic of the tapered microwire.

Figures 1(b) and 1(c) show the input facet and a schematic of the chalcogenide optical microwire used for observation of MI in the mid-IR. The microwire consists of an $As_2Se_3$ glass core surrounded by a polymer (PMMA) cladding, drawn using the technique described in Ref. [23]. The polymer and As$_2$Se$_3$ glass are heated and tapered together since they exhibit comparable glass-transition temperatures. The PMMA cladding makes the microwire resistant to mechanical strain and prevents optical interaction and damage with outside environment. Finite element simulations for the fundamental mode show that the fraction of light propagating in the polymer cladding is negligible (<0.5%) as the refractive index difference between As$_2$Se$_3$ and PMMA is very high ($\Delta n > 1.3$). The input core size in the untapered region is 16 µm and is reduced down to 3.6 µm in the 14-cm long uniform tapered region. The cutoff wavelength for the fundamental mode is 5.2 µm, but no evidence of multimode behavior was observed under our experimental conditions, which we attribute to weak intermodal coupling. The total insertion loss was also measured at 6 dB. As we discuss below, the zero-dispersion wavelength (ZDW) in the 3.6 µm region is ~2830 nm so that we pump in the normal GVD regime.

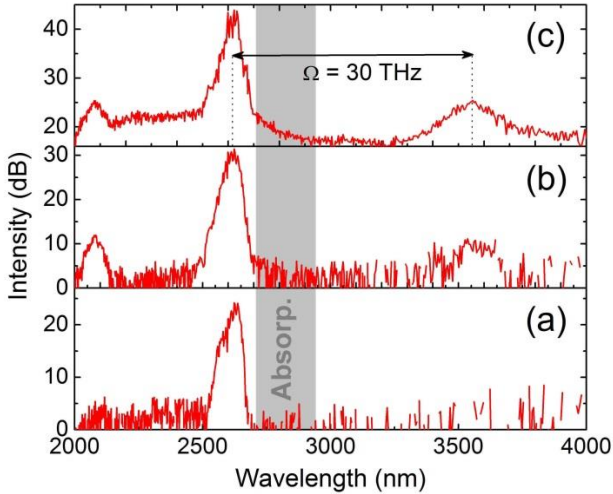

Fig. 2 Experimental observation of modulation instability when pumping at λ = 2620 nm in the normal dispersion regime. Input peak powers estimated at (a) 200 mW ; (b) 1W ; (c) 10 W.

Figure 2 shows the experimental spectra at the output of the microwire. Estimated input peak powers for these results were: (a) 200 mW ; (b) 1W ; (c) 10 W. We see the clear emergence of two widely spaced MI sidebands with increasing pump power. The short-wavelength sideband is centered at 2080 nm (540 nm from the pump) and the long-wavelength sideband is at 3555 nm (935 nm from the pump). These two sidebands are symmetric in frequency with a frequency shift $\Omega/2\pi = \pm 30$ THz. The shaded region in Fig. 2 indicates the OH absorption band of the As$_2$Se$_3$ microwire (located around 2.75 µm with a 200 nm width). We measured the loss (due to residual OH contamination) at ~13 dB m$^{-1}$ at the peak, but since the sidebands emerge from noise far from the absorption region, it has negligible influence on the MI frequency conversion process.

To interpret our experimental results quantitatively, we calculated the modal guidance and group-velocity dispersion (GVD) of the microwire and then performed numerical simulations of the pulse propagation. Fig. 3 shows the GVD parameter of the hybrid microwire for an increasing waist diameter for 3-16 µm. The GVD is numerically calculated from a step-index model for the hybrid As$_2$Se$_3$-PMMA microwire using known refractive index data.

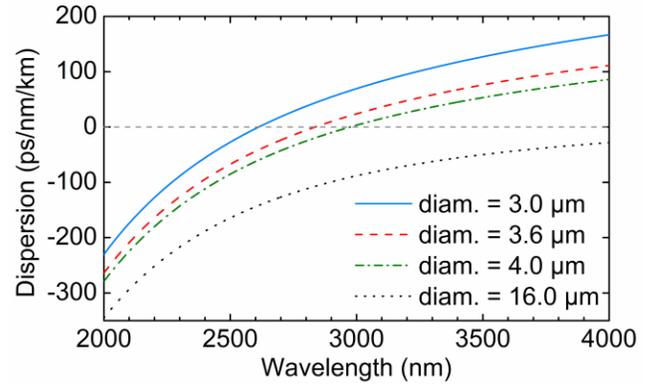

Fig. 3 GVD curves for selected microwire diameters.

For our microwire, the lead-in segment has diameter of 16 µm and length 2.8 cm. The 1.2 cm transition region reduces the diameter down to 3.6 µm and the diameter is maintained at this value for 14 cm. The output transition region and the lead-out section have 1.2 cm and 2.8 cm respectively. Figure 3 shows calculated dispersion for the fundamental mode for a selection of diameters of the fiber structure. For a pump at 2.6 µm, the fiber is normally dispersive in all segments of the fiber, but the nonlinearity is particularly enhanced in the central microwire region that is of interest in MI frequency conversion.

The dispersion calculations allow us to calculate the phasematching condition for normal dispersion MI using the standard stability analysis approach [10,14]. The dispersion relation is expressed as:

$$K = \frac{\beta_3 \Omega^3}{6} \pm \left[\left(\frac{\beta_2 \Omega^2}{2} + \frac{\beta_4 \Omega^4}{24}\right) \times \left(\frac{\beta_2 \Omega^2}{2} + \frac{\beta_4 \Omega^4}{24} + 2\gamma P\right)\right]^{1/2} \quad (1)$$

where the $\beta_i$s are the usual dispersion parameters, $\gamma$ is the nonlinear coefficient and P the pump peak power. Here, K is the wavenumber of the perturbation, and for a normal dispersion pump with $\beta_2 > 0$, gain (imaginary wavenumber) is observed around frequencies of

$$\Omega = \left(-\frac{12\beta_2}{\beta_4}\right)^{1/2} \quad (2).$$

At our pump wavelength of 2620 nm, calculations yield for the 3.6 µm microwire region dispersion parameters of: $\beta_2 = 2.9296 \times 10^{-2}$ ps$^2$ m$^{-1}$, $\beta_3 = 2.8428 \times 10^{-3}$ ps$^3$ m$^{-1}$, $\beta_4 = -9.89 \times 10^{-6}$ ps$^4$ m$^{-1}$. This predicts a MI gain frequency $\Omega/2\pi = 30$ THz, in agreement with the measured MI frequency.

The parametric gain as a function of pump wavelength for the 3.6 µm microwire is readily calculated from the imaginary part of Eq. (1), and the result is shown in Fig. 4. As the pump wavelength is varied over 2.6-2.9 µm, we clearly observe the transition between far detuned MI in the normal dispersion regime and the usual scalar MI in the anomalous dispersion regime. Note that the location of the MI peaks is in good agreement with the experimental observations of Fig. 2. Experiments tuning to shorter pump wavelengths showed qualitative evidence of shifting along the MI gain curve but with sidebands closer to the noise floor. Tuning the pump to longer wavelengths was not possible since the broadened pump then entered the absorption band, although the use of a narrowband pump (10's of ps or ns pulses) would be expected to yield broad tunability.

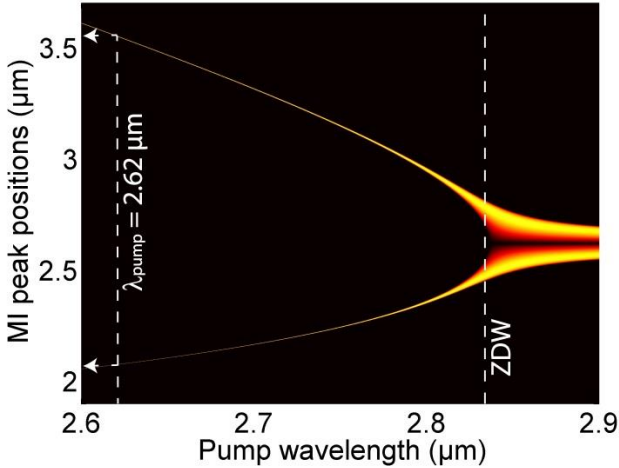

Fig. 4 Calculated MI normalized gain as a function of pump wavelength. The dashed lines indicate the positions of the MI sidebands at the pump wavelength and the zero-dispersion wavelength (ZDW). The sideband positions seen in experiment are also indicated.

We also performed numerical simulations of the MI sideband generation process solving the nonlinear Schrödinger equation (NLSE) [8]:

$$i\frac{\partial A(z,t)}{\partial z} = \frac{\beta_2}{2}\frac{\partial^2 A}{\partial t^2} + i\frac{\beta_3}{6}\frac{\partial^3 A}{\partial t^3} - \frac{\beta_4}{24}\frac{\partial^4 A}{\partial t^4} - \gamma|A|^2 A \quad (3)$$

using the usual split-step method and a stochastic noise model. We explicitly included variation in nonlinear and dispersion parameters according to the change in diameter in all segments of the chalcogenide waveguide, including the untapered sections, the transition regions and the uniform optical microwire. Random variations in diameter due e.g. to fabrication are negligible with the tapering process used [24]. We used the nonlinear response function $R(t) = (1-f_R)\delta(t) + f_R h_R(t)$ including the instantaneous Kerr contribution $\delta(t)$ and the delayed Raman contribution for As$_2$Se$_3$ given by $h_R(t) = [(\tau_1^2 + \tau_2^2)/(\tau_1 \tau_2^2)]\exp(-t/\tau_2)\sin(t/\tau_1)$ where $\tau_1 = 23.3$ fs, $\tau_2 = 230$ fs and $f_R = 0.1$ [2]. OH-absorption was modelled using a Gaussian line shape, but its influence on the output spectra was found to be weak. Neither two-photon absorption nor free-carrier absorption were taken into account as they become negligible for wavelengths longer than ~1770 nm [4].

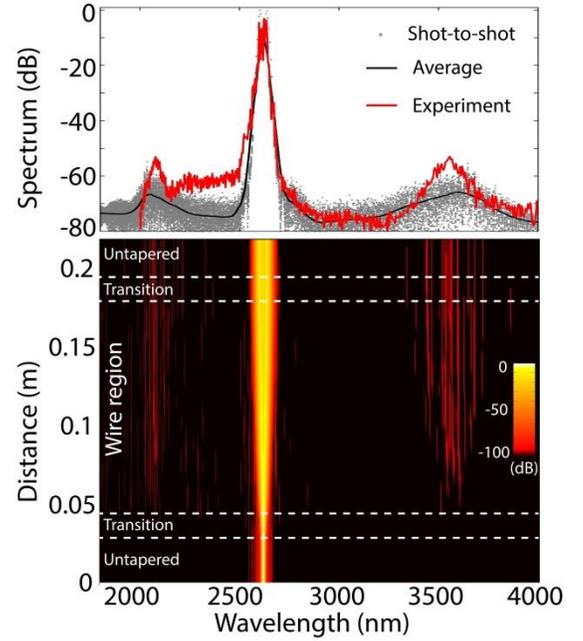

Fig. 5 *Top*: Simulation results showing mean spectrum at the microwire output (black) compared with experiment (red). Individual realizations from the ensemble (500 spectra) are also shown (gray dots). *Bottom*: Spectral evolution from one simulation simulated along the microwire length (total length of 22 cm) showing the dynamics of normal dispersion MI.

Figure 5 shows results obtained for the case of 10 W peak power, comparing experiment (red) with the average spectrum calculated from an ensemble of 500 realizations (black). Note that the individual shot-to-shot fluctuations in the simulations were significant as expected [25], and the figure also superposes the individual realizations (gray). The difference in sideband intensity between experiment and simulation is attributed to a higher level of background noise in experiment than modelled in the simulations. Indeed, simulations used a one-photon per mode quantum limited noise seed [10].

The bottom subplot in Fig. 5 shows the dynamical evolution in the different fiber segments. We see that the spectral generation of the MI components occurs only in the 3.6 µm diameter microwire region where the nonlinear coefficient $\gamma$ is the largest (at the pump wavelength, $\gamma \sim 7$ W$^{-1}$.m$^{-1}$ in the wire region whereas $\gamma \sim 0.3$ W$^{-1}$.m$^{-1}$ in the untapered region). Additional simulations showed that the effect of the transition region is negligible, and the influence of the Raman contribution is low, with no effect on the MI peak position or shape. It is however responsible for the apparition of a low-intensity peak on the long-wavelength side of the pump.

In conclusion, we have demonstrated a mid-infrared parametric wavelength converter, realized by pumping a highly nonlinear chalcogenide tapered fiber with a femtosecond optical parametric oscillator. Far detuned frequency conversion with 1.5 µm wavelength spacing has been achieved in the mid-IR using scalar modulation instability in the normal dispersion regime. The 30 THz frequency shifts observed are the largest reported using normal dispersion pumped scalar MI in a single-pass configuration. Experimental observations were verified using numerical simulations based on a stochastic NLSE model. These results show the potential of chalcogenide microwires for far-detuned mid-infrared frequency conversion with potential applications including entangled photon pair generation, absorption spectroscopy and chemical sensing.

This work was supported by the European Research Council (ERC) Advanced Grant ERC-2011-AdG-290562 MULTIWAVE, the Agence Nationale de la Recherche (ANR OPTIROC), the Air Force Office of Scientific Research grant number FA8655-13-1-2137, and the Natural Science and Engineering Research council of Canada (NSERC).